\documentclass[11pt,twoside]{article}
\usepackage{asp2004}
\usepackage{psfig}
\usepackage{epsf}
\usepackage{graphics}
\usepackage{lscape}
\def\edcomment#1{\iffalse\marginpar{\raggedright\sl#1\/}\else\relax\fi}
\reversemarginpar

\begin{document}
\title{Large scale correlations of quasar polarisation vectors: 
Hints of extreme scale structures?}
\author{R\'emi~A. Cabanac$^1$, Damien Hutsem\'ekers$^2$, Dominique Sluse$^2$,
Herv\'e Lamy$^3$}
\affil{$^1$ Canada-France-Hawaii Telescope, 65-1238 Mamalahoa Highway, Kamuela, HI 96743, USA.\\
$^2$Institut d'Astrophysique et de G\'eophysique, ULg, All\'ee du 6 Ao\^ut 17, B5C, 4000 Sart Tilman (Li\`ege), Belgium.\\
$^3$BIRA-IASB, Avenue Circulaire 3, 1180 Bruxelles, Belgium.\\ }

\begin{abstract}
A survey measuring quasar polarization vectors has been started in
two regions towards the North and South Galactic Poles. Here, We review the
discovery of significant correlations of orientations of polarization vectors
over huge angular distances. We report new results including a larger sample
of the quasars confirming the existence of coherent orientations at redshifts
$z>1$.
\end{abstract}
\thispagestyle{plain}

\section{Introduction}

Large scale alignments of quasar polarization vectors were uncovered
by Hutsem\'ekers (1998), looking at a sample of 170 QSOs selected from the
litterature and confirmed later on a larger sample 
(Hutsem\'ekers \& Lamy 2001). The departure to random orientations was found at
significance levels small enough to merit deeper investigations.  
Moreover, these alignments seemed to come from high
redshift regions, implying that the underlying mechanism might cover
physical distances of Giga parsecs. A large survey of linear polarization 
was then started, with the long-term goal to characterize better the 
polarization properties of quasars, and a short-term goal to investigate 
the reality of the alignments. This work gives a preliminary
analysis of the alignment effect for a total sample of 355 quasars, 
comprising new polarization measurements from observing runs between 
2001-2003 and a new comprehensive compilation from the litterature.

\section{Sample}

The sample was chosen from V\'eron-Cetty \& V\'eron (2000) and the Sloan 
Digital Sky Survey. Among possible targets a preference was given to bright 
extragalactic sources having a higher probability to show stronger
 polarization: BAL quasars, red quasars. The selection criteria of the 
sample are detailed elsewhere (Hutsem\'ekers 1998, Hutsem\'ekers et al. 2001, 
2004; Sluse et al. 2004).
In order to avoid possible contaminations from the interstellar medium 
of the Galaxy, only the objects at galactic latitude $|b_{gal}|>30^\circ$
were selected. Above this galactic latitude the interstellar 
polarization is smaller (Heiles 2000).

The observations were carried out at ESO La Silla 3.6m EFOSC2 (2001, 2002, 
2003) and Paranal VLT/FORS1 (2003), in broad-band $V$ filter, using a
Wollaston prism and 4 positions of the half-wave plate to derive Stokes 
parameters $Q$ and $U$ (Hutsem\'ekers et al. 2004; Sluse et al. 2004).
The photometry was done using the procedure defined by Lamy \& Hutsem\'ekers 
(1999). Photon errors on $Q$ and $U$ are $\sigma_Q\simeq\sigma_U = 0.16\,\%$. 
The foreground polarization (coming from both the instrument
and the interstellar medium) was computed from the observed field stars 
and subtracted from the quasar polarization. Because this was not possible 
on every frame, we used Stokes parameters averaged from the field stars 
observed during the same run. This average polarization is always small : 
$\overline{q_*}\simeq\overline{u_*}\simeq 0.1\pm0.2\,\%$. 
Final errors were conservatively derived by quadratically adding photon 
and foreground errors yielding a total error on the QSO polarization degree
ca. 0.2-0.3 \%. Finally, because the polarization is a positive quantity only, 
it is debiased according to the Wardle \& Kronberg method (1974). 
To insure a robust measurement of the polarization direction we include only
the objects having a polarization degree of $P>0.6\,\%$ and a maximum error
in the polarization position angle of $\sigma_\theta<14\,^\circ$.

The final sample of polarized quasars comprises 195 quasars in the North 
Galactic Pole region (NGP) and 160 quasars in the South Galactic Pole 
region (SGP). Figure 1 shows an Aitoff projection of the full sample
superimposed on the galactic star polarization map (Heiles 2000).
\begin{figure}
\plotone{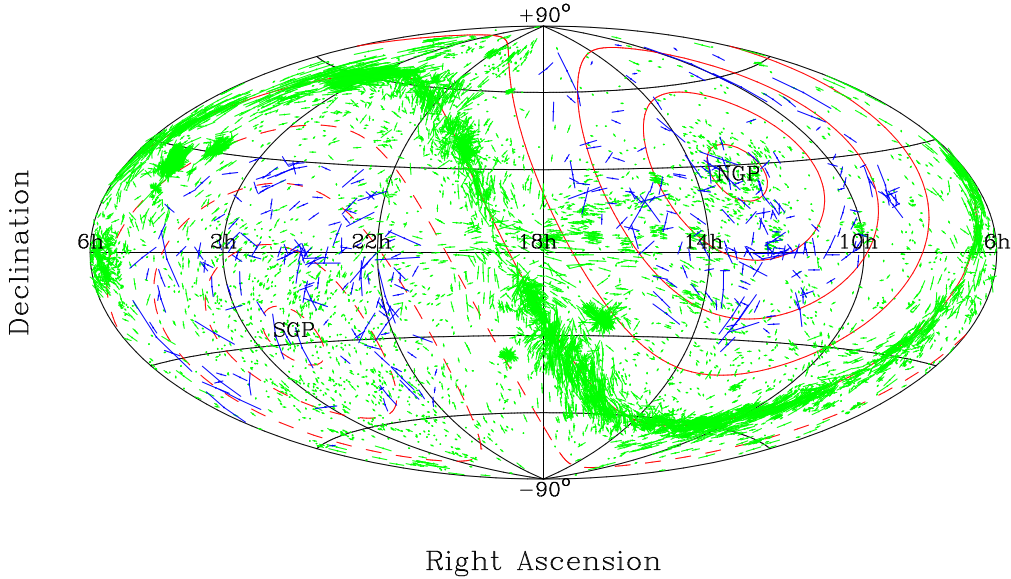}
\caption{Aitoff projection of the full sample of polarized quasars 
(black segments) and galactic stars (grey segments; Heiles 2000). 
Contours are 20$^\circ$-step galactic parallels (solid lines towards NGP; dashed lines towards SGP). In polar regions the quasar polarization is on average 
5 times stronger than the star polarization.
}
\end{figure}

Our sample spans the redshift range $0<z<3$ homogeneously, with a marginally 
better sampling in the range $0<z<1$, mainly due to our bias towards brighter
quasars.  Figure 2 shows a plot of the quasar sample polarization degrees 
against redshift (left panel). The slight enhancement 
of highly polarized quasars at lower redshift is attributed to the fact that 
more radio-loud quasars were observed at low redshifts. Overall the quasar 
polarization degrees do not correlate with redshift which is consistent 
with previous studies (Berriman et al. 1990). 
In Fig.\,2 right panel, the histograms 
of the linear polarization degree of the quasar sample (black line) and 
of the star sample (Heiles 2000) selected over the same region (grey line) 
are superimposed.
The star polarization histogram shows much smaller polarization degrees 
(94 out of 1996 stars have a polarization degree of 0.5\% or higher)
arguing in favor of minor contaminations from the interstellar medium 
on the quasar sample. 
\begin{figure}
\plotone{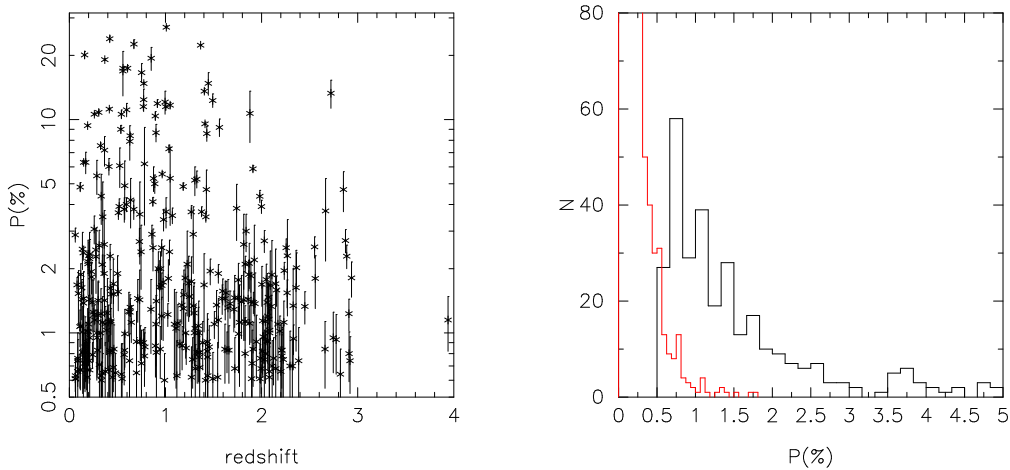}
\caption{Left: The polarization degrees vs redshift do not show 
correlations. The slight enhancement of highly polarized quasars at lower
redshift is attributed to an observational bias. Right: The polarization 
histogram of the quasar sample (black) is compared to the star polarization 
histogram (grey; Heiles 2000) selected from the same regions. 
The star sample shows much smaller polarization degrees.
}
\end{figure}

\section{Statistical significance of alignments}

In order to detect and assess the significance of the alignments over the 
complete 3D sample, we used two statistical tests designed for circular
data. The detailed procedure is described in Hutsem\'ekers (1998). The
basic idea is to compute for each quasar a statistics $S$ taking into account 
the compactness of a group of $n_v$ neighbors (in redshift space) and the 
dispersion of their polarization direction. The more compact and aligned a 
group of $n_v$ neighbors, the smaller the statistics $S$. 
The second statistical test is the Andrew-Wasserman test designed by 
Bietenholz (1986). 
Both statistics can be calculated for the entire sample and 
the significance of a departure from homogeneity is assessed with Monte-Carlo 
reshuffling of the polarization direction over 3-D positions. 
The significance level is then computed as the probability that a random 
configuration has a smaller ($S$ test) or higher statistics (Andrew-Wasserman
test) than the observed one. 
Figure~3 Right frame shows the logarithm of the 
significance level of the Andrew-Wasserman test versus the number of 
neighbors $n_v$ for three samples of quasars. It clearly shows that, 
as the size of the sample increases, the alignments are harder and harder to 
produce from random distributions for any value of $n_v$.
Fig.~3 Left frame show the polarization map of the high redshift NGP sample 
(top) and the low redshift NGP sample (bottom). The length of the bars is
proportional to $\sqrt{P}$. Objects of both high and low polarization degrees
participate to the same alignments.
An unambiguous alignment effect of the polarization direction is visible at
high redshift, whereas another alignment emerges at low redshift in a different 
direction.
\begin{figure}
\plotfiddle{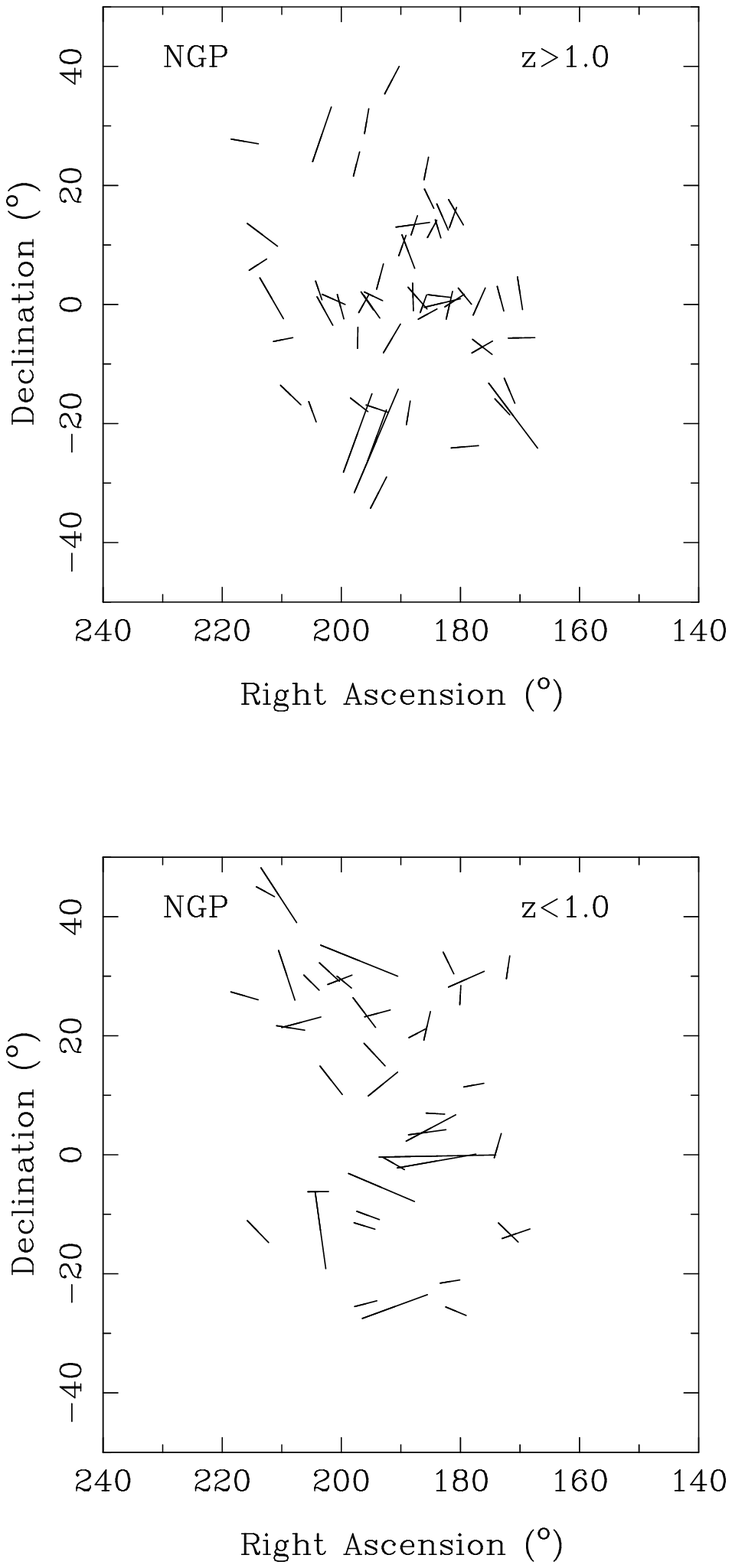}{8cm}{0}{45}{45}{-180}{-40}
\plotfiddle{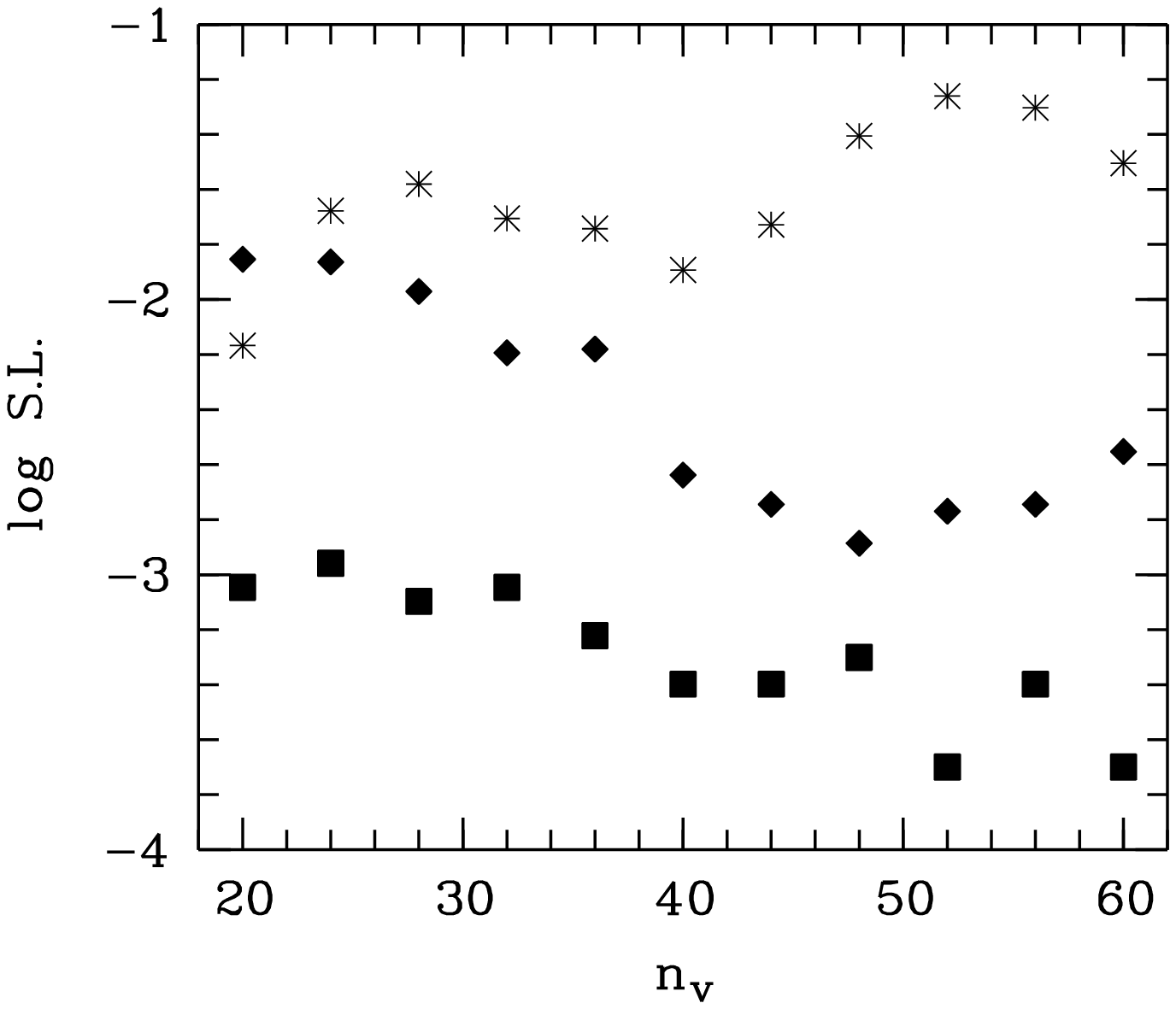}{0cm}{0}{50}{50}{-50}{-40}
\caption{Right: Logarithm of the significance level of the alignments versus 
the number of neighbors for the original sample of 170 quasars (stars), an
intermediate sample of 213 quasars from Hutsem\'ekers \& Lamy (2001; diamonds),
and the present sample 355 quasars (squares). It is more and more difficult 
to produce the observed alignments from random distributions. 
Left: Maps of the region NGP in a high (top) and a low (bottom) redshift range.
The length of the bars is proportional to $\sqrt{P}$. 
Low and high redshift alignment directions are distinct.}
\end{figure}

Our statistical tests are not invariant under polar coordinates. 
A way to avoid coordinate dependence is to parallel transport the
angles along great circle on the celestial sphere prior to computing the
statistics. This was done by Jain et al. (2004). They confirm 
the significance of the alignments of the 213-quasar
sample over very large regions of the NGP and SGP.

\section{Discussion}

The instrumental polarization can be discarded as a source of contamination 
because, it is very small on EFOSC2 ($\sim$0.1\%) and because the polarization 
degree and angle of quasars observed on different instruments are
consistent within the errors. The contamination from the Galaxy is
not a dominant component. Indeed, the interstellar polarization is usually 
much smaller than the polarization of quasars, and the interstellar 
polarization angles do not follow the quasar alignment directions.
Moreover it is difficult to explain the redshift dependence of the alignment 
directions assuming a foreground interstellar screen.
If we accept the fact that the alignments are intrinsic to the sources, we
have to face correlated polarizations over extreme scales of Giga parsecs.
Exotic sources of polarization on cosmological scales could be invoked,
such as pervading exotic particles (Harari \& Sikivie 1992; Jain et al. 2004) 
or an intrinsic alignment of the axes of quasar central engines. 
Future theoretical developments need to take these observed correlations into 
account. 

The ongoing survey is paramount to characterize the alignments of 
quasar polarization directions accross the sky and correlate them with 
the other large-scale dataset (CMB, galaxy surveys, ...). It will also 
allow us to deepen our knowledge of the linear polarization properties of 
quasars, and to understand the different source of linear 
polarization between the different species. 
We definitely need to increase the sample of polarized quasars 
from a few hundreds to a few thousands over the next decade to progress 
significantly in the field.


\begin{references}
\reference Berriman, G., Schmidt, G.~D., West, S.~C., \& Stockman, H.S., \apjs, 1990, {\bf 74}, 869.
\reference Bietenholz M.~F., 1986, \aj, {\bf 91}, 1249.
\reference  Harari, D., \& Sikivie, P. 1992, Phys. Lett. B, {\bf 289}, 67
\reference Heiles, C., 2000, \aj, {\bf 119}, 923. 
\reference Hutsem\'ekers, D., Sluse, D., Cabanac, R., Lamy, H., Quintana, H., in prep.
\reference Sluse, D.,  Hutsem\'ekers, D., Lamy, H., Cabanac, R., in prep. 
\reference Hutsem\'ekers, D., 1998, A\&A, {\bf 332}, 410.
\reference Hutsem\'ekers, D., \& Lamy, H., 2001, A\&A, {\bf 367}, 381.  
\reference Jain, P., Narain, G., \& Sarala, S., \mnras, {\bf 347}, 394.
\reference Lamy, H. \& Hutsem\'ekers, D., 1999, Msngr., {\bf 96}, 25. (Erratum: The Messenger 97, 23)
\reference Wardle, J.~F.~C. \& Kronberg, P.~P., 1974, \apj, {\bf 194}, 249.  
\end{references}
\end{document}